# Teaching quantum mechanics to over 14,000 nonscientists

J. K. Freericks[1], L. B. Vieira[2], D. Cutler[1], and A. Kruse[1]

[1]Georgetown University, 37th and O Sts. NW, Washington, DC 20057 USA

[2]Universidade Federal de Minas Gerais, 6627 Av. Antônio Carlos, Belo Horizonte, MG, Brazil 31270-901

We describe a new style of MOOC designed to engage students with an immersive multimedia environment including text, images, video lectures, computer-based simulations, animations, and tutorials. *Quantum Mechanics for Everyone*[1] is currently running on EdX and has been successful by a number of different measures. It illustrates both how one can teach complex quantum phenomena to nonscientists and how one can develop high quality interactive computer simulations that engage the students and can be widely deployed.

We have been told a lie. We promote the lie. We internalize the lie. And we don't even know where the lie originated. The lie is that *one needs to have a sophisticated math background and significant physics background in order to understand complex quantum phenomena*. This article dispels the lie. It shows you how to bring the most exciting and counterintuitive physics concepts to the forefront of the curriculum. While our initial emphasis has been to reach nonscientists via a MOOC format, many of the techniques we employ can be rolled out in a classroom setting and included into introductory curricula. It can be done in a meaningful way that is honest to the science. It can be done without requiring sophisticated math.

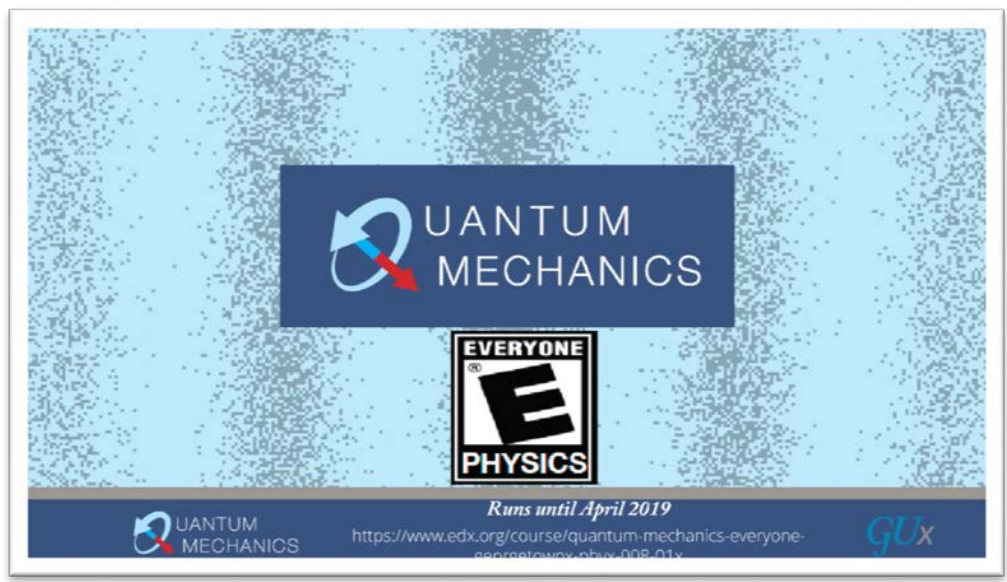

*Fig. 1. Logo for the MOOC..*

But don't just take our word for it. The initial idea comes from Richard Feynman, and is eloquently expounded in his book entitled *QED: The Strange Theory of Light and Matter*[2]. There Feynman illustrates how the path-integral method for quantum mechanics can be developed by just drawing arrows on a piece of paper! He discusses much of the quantum mechanics of light, ranging from everyday phenomena, like partial reflection, mirrors, and lenses, to more exotic phenomena like the two-slit experiment. The idea was further developed by Daniel Styer, in *The Strange World of Quantum Mechanics*[3], who applied the Feynman methodology to Stern-Gerlach experiments which were then modified to allow him to describe the two-slit experiment, Wheeler's delayed choice experiment, the Einstein-Podolsky-Rosen paradox, and Bell's theorem experiments. In our MOOC, which is running on EdX until the Spring of 2019, we combine all of these different ideas and more into a multimedia educational experience for the students (see Fig. 1).

Soon after Feynman's book came out, Edward Taylor realized that animations of the material in the book would enhance the learning process[4]. He developed a series of computer-based tutorials that employed the cT programming environment to run these animations. These computer tutorials were further developed by Hanc and Tuleja who rewrote them in Java and provided an enhanced learner experience[5]. Unfortunately, neither of these approaches work well for web delivery of materials, where Java programs suffer from numerous security concerns and there is no web interface to run cT programs. As we worked on developing the materials for our MOOC, we focused on creating platform independent computer tutorials which would run off of JavaScript and be deployed via iframes on a web browser. This industry-standard approach allows for the computer simulations to have the widest reach and the highest impact.

Because modern browsers and JavaScript have numerous high quality graphics libraries and 3D imaging tools, we could produce animations that were engaging and realistic using modest resources. The end result is a professional, game-quality, series of animations that work on a wide range of different platforms, including computers, tablets and phones. We incorporate dozens of these simulations within the MOOC. As an added bonus, we employed undergraduates to program these simulations, leading to unique senior research projects which combined physics, art, and computer science.

*Quantum Mechanics for Everyone* delivers a self-contained treatment of quantum theory that is accessible to all with a modest knowledge of high-school-level algebra. The initial results, after five months of the 24 month run, are impressive–we have over 14,000 learners enrolled, the course is the highest rated quantum MOOC on coursetalk.com, and over 145 students have already completed the final.

The class is organized into four modules–each intended to be covered in one week. The first two modules are heavily influenced by Styer's book. The second two, by Feynman's. The first module covers an introduction to quantum mechanics, where we describe classical expectations for a Stern-Gerlach experiment and then illustrate what happens when it is performed on quantum particles. This leads into a need to describe systems probabilistically and allows us to introduce a unit on probability theory to ensure students understand how to determine the probabilities for different events to occur. The second module, on advanced quantum mechanics with spin, develops the concept of the Stern-Gerlach analyzer loop, which involves two inverted Stern-Gerlach analyzers hooked back-to-back to allow for quantum interference effects to be carefully examined. Armed with this devices, we introduce an analog of the two-slit experiment, Wheeler's delayed choice variant and the quantum eraser, the EPR experiment, and a Bell-state experiment, which allows us to rule out hidden variables as a viable description for quantum phenomena. The module ends with a discussion of the technology behind magnetic resonance imaging machines. The third module involves the quantum mechanics of light, where we describe partial reflection, the single-slit experiment, the two-slit experiment (both watched and unwatched), multi-slit experiments, mirrors, and lenses. This module fully develops the quantum rules by employing Feynman's model for path integrals. The fourth module concludes the course and covers the phenomena of quantum seeing in the dark (or interaction-free measurements) by first illustrating how they would work with a two-slit experiment. Then, describing the Mach-Zehnder interferometer, and how this vastly improves the efficiency of these kinds of experiments. This is followed by a treatment of the polarization of light and the quantum-Zeno effect, which is finally incorporated within a Mach-Zender interferometer to conduct the quantum seeing in the dark experiment (which involves seeing an object *without* shining light on it). The unit concludes by describing the properties of bosons and indistinguishability as evidenced by photon bunching in the Hong-Ou-Mandel experiment. For further details, please visit the EdX course site at https://www.edx.org/course/quantum-mechanics-everyone-georgetownx-phyx-008-01x.

This journey through quantum mechanics takes the learner on a trip that visits increasingly complex and abstract phenomena, but at a level that is easy to follow and understand. The journey is made possible only through the use of sophisticated computer-based simulations and animations which engage the students but also allow them to visualize the abstract phenomena and make it real. Furthermore, it performs the complex quantum calculations for the students, so they can focus on understanding the methodology, but not get caught up in the minutiae of the details.

Since the development of the computer tutorials is so germane to the course, we take a moment to describe some of the issues one must deal with if you wish to embark on a similar endeavor for another topic (of course, if you wish to use any or all of our tutorials, they are freely available for download under the LGPL 2.1 license at our GitHub repository https://github.com/quantum-mechanics-for-everyone/simulations). As previously mentioned, it is vital that the tutorials rely only on standard and widely deployable features of major web browsers. The maturity of JavaScript and WebGL in modern browsers (desktop and mobile) is a key factor for our success. Modern browsers are powerful multimedia platforms and so are an ideal medium for interactive educational content that can be widely deployable–now freed from third-party software such as Java or Flash.

In order to have a flexible system to develop the tutorials, they are envisioned as simulations based on a custom-built framework featuring visually-distinct modular devices, with inputs and outputs, which may be attached to one another, rotated and labeled at will. This core simulation engine allows us to quickly develop many different experimental setups with great flexibility (See Fig. 2).

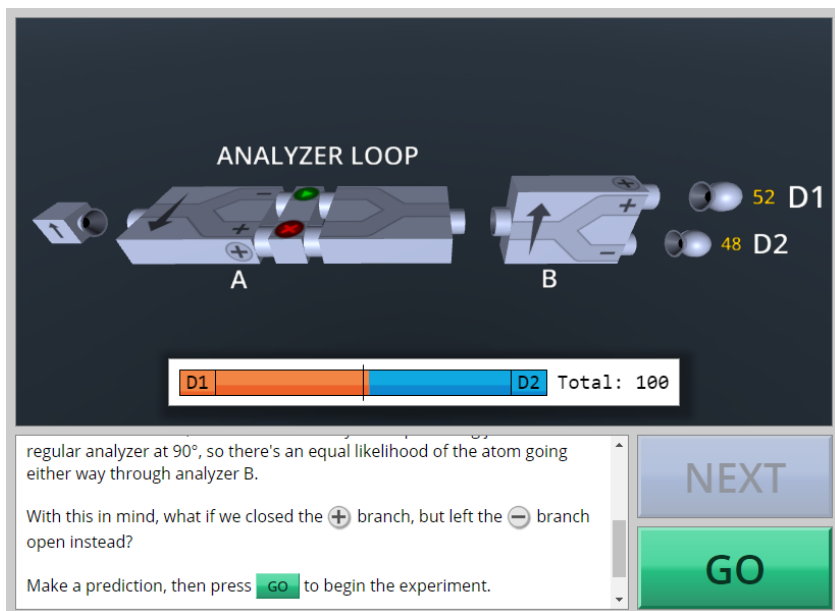

*Fig. 2. An example of a virtual experimental setup. On the left, a spin-up "atom source" releases atoms with a known spin state (+z, vertical). This atom passes through a Stern-Gerlach analyzer connected to a partially-open "gate" object, which in turn is connected to a "quantum eraser". These three objects combine to form an "analyzer loop". The atom then goes through a second Stern-Gerlach analyzer before reaching detectors D1 or D2. Below, a color-coded proportional bar chart displays the result of the experiment.*

In our engine, particles are released by sources and pass through these devices until they reach detectors, at which point a measurement result is tallied up and displayed in bar charts. Specific care is taken to avoid classical vs quantum misconceptions with this visual approach: quantum behavior always occurs inside the devices, hidden from view. A textbox accompanies all simulations and gently guides the learner through the various steps, providing explanations and posing questions to enhance the experience.

A probabilistic model is automatically computed by the engine to reproduce the results expected of these "virtual experiments", which allow us to abstract the software development away from the technical details in each tutorial and focus on the narrative and presentation.

Each object is given a distinct shape and name (e.g. "atom source", "Stern-Gerlach analyzer", "detector") as to be immediately recognized, and is introduced and explained in detail and in terms of previously established concepts. Care was taken to ensure possible questions about their behavior were addressed early on ("What happens if we flip this upside down, or attach it to the other output?"). This way, students are quickly familiarized with the full range of behaviors and usage. Questions to be addressed in later experiments are also mentioned earlier in order to prevent leaving the more curious and attentive students empty-handed.

An important aspect of the probabilistic model for the simulations is that it allows us to visually emphasize the probabilistic nature of quantum mechanics. Experiments always consist of hundreds of consecutive trials until conclusions can be drawn from the results. They are displayed using intuitive visuals like color-coded proportional bar charts, which give an immediate intuition for probabilities between two events. As results start to accumulate, the student can almost "feel" the results converging to the predicted value, despite the random fluctuations. This greatly enhances the satisfaction of running the simulations, making them more concrete and tangible, as opposed to simply reporting the final result (see Fig. 3)

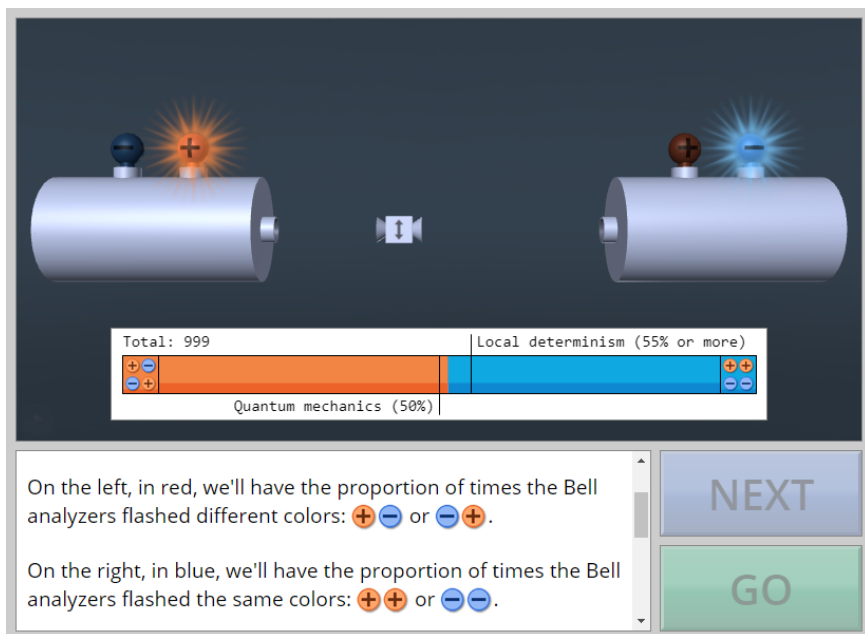

Fig. 3. Bell experiment using two "Bell analyzers" and a source of entangled atom pairs. Each Bell analyzer contains a three-positioned Stern-Gerlach analyzer and two detectors inside, as previously assembled in an earlier tutorial.

Each experiment is built piece by piece based on previous ones. Once set up, we discuss what we have found so far in previous tutorials, then pose questions about the new experiment. Before any experiment begins, we encourage (but do not force) students to make predictions of the results. This makes the simulations more engaging and encourages critical thinking. After each experiment, we ask if the results were what the student expected.

Sometimes, we deliberately play with classical intuition by presenting an unexpected quantum result, followed by a discussion of the incorrect assumptions. In this way, we create a narrative that deconstructs classical ideas in favor of quantum ones. More complex experiments are first simulated step by step, pausing at important points, so aspects of each device can be explained in more detail, before being run a series of times in succession.

Finally, in order to make the simulations visually engaging, we also included smooth movements and transitions when moving the camera or devices in the experiments. Devices are always brought from outside or taken from inside the experiment and moved out, and the camera position and angle is chosen to accommodate the new experimental setup.

What advice do we have for other MOOC developers: (i) do not underestimate the time it takes to develop materials---this four week course took three years to complete; (ii) be sure to test your materials multiple times prior to release---watch every video, read every transcript, try every problem, select alpha-testers from your target audience to give learner-appropriate feedback; (iii) be prepared to correct your course as errors are found---no course is error free and students will find many during the initial course roll-out; and (iv) be responsive on the discussion boards---nothing is more annoying than having limited staff participation to ensure the course is running smoothly and that problems are being fixed.

Finally, what is the implication for education? As we often bemoan the shrinking numbers of physics majors (in spite of some recent reinvigoration of numbers nationwide), we should ask ourselves how can we best excite students to want to learn more physics? While some may say the traditional methods with blocks and pulleys and charges and fields is the way to go, we feel that the physics community should include some quantum mechanics into the curriculum earlier than normally done to benefit all. Now there is a roadmap for how this goal can be achieved!

You might also ask is this all that can be taught in quantum mechanics without requiring a sophisticated mathematical edifice? The answer here is surprisingly also no! In fact, one of us (JKF) is already embarked on writing a book entitled *Quantum Mechanics Without Calculus*, which will teach many elements of the standard undergraduate and graduate quantum curriculum and even move into research active fields, requiring math that only involves high-school-level algebra. While students will need to be *masters* of high-school-level algebra to finish such a book, one can cover nearly all of quantum mechanics by employing operator based methods and not ever needing to calculate a derivative or an integral! Perhaps the release of this book will finally free us from the quantum lie that one needs calculus and differential equations to be able to truly understand the subject. It turns out that one only needs algebra skills and an ability to reason abstractly. It is time we bring this excitement to our students as soon as we can. We hope this MOOC will only be the first step on this journey. And our students will all be the better for it.

Acknowledgments: This work was supported by the National Science Foundation under grants numbered PHY-1314295 and PHY-1620555, by an ITEL grant from Georgetown University, and by the McDevitt bequest from Georgetown University.